\documentclass[a4paper,11pt]{article}
\usepackage{amsmath}

\usepackage[utf8]{inputenc}
\usepackage[DIV=12]{typearea}\usepackage{ragged2e}
\usepackage{calligra,amsmath,amsfonts,bbm,mathrsfs,amssymb}
\usepackage{slashed,cancel}
\usepackage{cite}
\usepackage{bm} 
\usepackage{hyperref}
\usepackage[dvipsnames]{xcolor}
\usepackage{colortbl}
\usepackage{graphicx}
\usepackage{cleveref}
\usepackage{bm}
\usepackage{mathdots}
\usepackage{color}
\usepackage{dsfont}
\usepackage{slashed}
\usepackage{physics}
\usepackage{xcolor}
\usepackage{subfigure}
\usepackage[normalem]{ulem}
\usepackage{comment}
\usepackage{dsfont}
\usepackage{lipsum}
\usepackage{mathtools}
\usepackage{url}
\usepackage{dcolumn}
\usepackage{booktabs}

\newcommand{\email}[1]{\href{mailto:#1}{\tt #1}}

\def\be{\begin{equation}}
\def\ee{\end{equation}}
\def\bea{\begin{eqnarray}}
\def\eea{\end{eqnarray}}

\begin{document}
\renewcommand*{\thefootnote}{\fnsymbol{footnote}}

\begin{titlepage}
\vspace*{4cm}

\begin{center}
 {\Large\bf Radiative Quarkonium decays into an Invisible ALP}
\centering
\vspace{0.8cm}

{\bf Luca Di Luzio$^{a}$\footnote{\email{luca.diluzio@pd.infn.it}}, Alfredo Walter Mario Guerrera$^{a,b,c}$\footnote{\email{alfredowaltermario.guerrera@st.com}}
, \\ Xavier Ponce D\'iaz$^{a,b}$\footnote{\email{xavier.poncediaz@pd.infn.it}}, \bf Stefano Rigolin$^{a,b}$\footnote{\email{stefano.rigolin@pd.infn.it}}}\\[7mm]

$^a$~{\it Istituto Nazionale di Fisica Nucleare (INFN), Sezione di Padova, \\
Via F. Marzolo 8, 35131 Padova, Italy}\\[1mm]
$^b$~{\it Dipartimento di Fisica e Astronomia ``G.~Galilei'', Universit\`a di Padova,
 \\ Via F. Marzolo 8, 35131 Padova, Italy}\\[1mm]
$^c$~{\it STMicroelectronics, Stradale Primosole 50, 95121 Catania, Italy}\\[1mm]

\vspace{0.3cm}
\begin{abstract}
In this talk the ALP production from radiative quarkonium decays is presented. To this purpose, the relevant 
cross-section is computed from a $d=5$ effective Lagrangian containing simultaneous ALP couplings to $b(c)$-quarks 
and photons. The interplay between resonant and non-resonant contributions is shown to be relevant for experiments 
operating at $\sqrt{s}=m_{\Upsilon(nS)}$, with $n=1,2,3$, while the non-resonant contribution dominates at the 
$\Upsilon(4S)$ resonance. These effects imply that the experimental searches performed at with different experimental 
setups are sensitive to complementary combinations of ALP couplings. To illustrate these results, constraints 
from existing BaBar, Belle and Belle II data on ALPs decaying into invisible final states are derived. Finally, 
constrains from the recent BESIII measurements of the $J/\Psi \to \gamma \,a$ decay rate are also included for 
comparison.
\end{abstract}

\thispagestyle{empty}

\end{center}
\end{titlepage}

\setcounter{footnote}{0}

% \title{Radiative Quarkonium decays into an Invisible ALP}
% 
% \author{L. Di Luzio$^a$, X. Ponce D\'iaz$^{a,b}$, A.W.M. Guerrera$^{a,b,c}$ and S. Rigolin$^*$ $^{a,b}$} 
% 
% \address{\vskip +0.2cm $^a$INFN Padova, Via Marzolo 8, Padova, Italy }
% \address{\vskip -0.2cm $^b$Department of Physics and Astronomy University of Padova , Via Marzolo 8, Padova, Italy }
% \address{\vskip -0.2cm $^{c}$STMicroelectronics, Stradale Primosole 50, 95121 Catania, Italy }
% 
% 
% \maketitle\abstracts{
% In this talk the ALP production from radiative quarkonium decays is presented. To this purpose, the relevant 
% cross-section is computed from a $d=5$ effective Lagrangian containing simultaneous ALP couplings to $b(c)$-quarks 
% and photons. The interplay between resonant and non-resonant contributions is shown to be relevant for experiments 
% operating at $\sqrt{s}=m_{\Upsilon(nS)}$, with $n=1,2,3$, while the non-resonant contribution dominates at the 
% $\Upsilon(4S)$ resonance. These effects imply that the experimental searches performed at with different experimental 
% setups are sensitive to complementary combinations of ALP couplings. To illustrate these results, constraints 
% from existing BaBar, Belle and Belle II data on ALPs decaying into invisible final states are derived. Finally, 
% constrains from the recent BESIII measurements of the $J/\Psi \to \gamma \,a$ decay rate are also included for 
% comparison.}

\section{Introduction}

Light pseudoscalar particles naturally arise in many extensions of the Standard Model (SM), including the ones 
endowed with an approximate global symmetry spontaneously broken at a given scale, $f_a$. Sharing a common nature 
with the QCD axion~\cite{Peccei:1977hh,Wilczek:1977pj,Weinberg:1977ma}, (pseudo) Nambu-Goldstone bosons are 
generically referred to as Axion-Like Particles (ALPs). The ALP mass $m_a$ can, in general, be much lighter 
than the symmetry breaking scale $f_a$, as it is paradigmatically exemplified in the KSVZ and DFSZ invisible 
axion models~\cite{Kim:1979if,Shifman:1979if,Zhitnitsky:1980tq,Dine:1981rt}. Therefore, it may be not 
inconceivable that the first hint of new physics at (or above) the TeV scale could be the discovery of a 
light pseudoscalar state. 

In this talk the existing BaBar and Belle flavor-conserving constraints on ALPs are carefully examined. In 
fact, the resonant contributions to the ALP production, via the $e^+ e^- \to \Upsilon(nS)\to a\gamma$ process, have 
been previously overlooked. As will be shown, these effects can induce numerically significant corrections to 
experimental searches performed at $\sqrt{s}=m_{\Upsilon(nS)}$, with $n=1,2,3$. A detailed analysis of these effects 
can be found in~\cite{Merlo:2019anv}. In order to assess the limits on ALP couplings, one should specify not only 
the ALP production mechanism, but also its decay products. Here, it will be assumed that the ALP does not decay 
into visible particles. Such a scenario can be easily achieved by assuming a sufficiently large ALP coupling to 
a stable dark sector, as motivated by several dark matter models. The conclusions related to ALP production are, 
however, general and they can also be applied to the reinterpretation of experimental searches with visible decays 
in the detector~\cite{DiLuzio:2024jip}.

\subsection{ALP effective Lagrangian}
\label{subsec:ALPEFF}

The dimension-five effective Lagrangian describing ALP interactions, above the electroweak symmetry breaking scale, 
can be generically written as~\cite{Brivio:2017ije}
%
%%%%%%%%%%%%%
\begin{align}
\begin{split}
\delta\mathcal{L}_{\mathrm{eff}}  =  \dfrac{1}{2} (\partial^\mu a) \,(\partial_\mu a) - \dfrac{m_a^2}{2}a^2 &- 
  \dfrac{c_{aBB}}{4} \dfrac{a}{f_a}\, B^{\mu\nu}\widetilde{B}_{\mu\nu}  
  - \dfrac{c_{aWW}}{4} \dfrac{a}{f_a}\, W^{\mu\nu}\widetilde{W}_{\mu\nu} \\
  & - \dfrac{c_{agg}}{4} \dfrac{a}{f_a}\, G_{a}^{\mu\nu}
      \widetilde{G}^a_{\mu\nu} - \dfrac{\partial_\mu a}{2 f_a} \sum_f c_{aff}\, \overline{f}\gamma^\mu \gamma_5 f \,,
  \label{eq:leffalp0}
\end{split}
\end{align}
%%%%%%%%%%%%%
%
where $\widetilde{V}^{\mu\nu}=\frac{1}{2} \,\varepsilon^{\mu\nu\alpha\beta} V_{\alpha \beta}$, $c_{aff}$ and $c_{aVV}$ 
denote the ALP couplings to fermions and to the SM gauge bosons,~$V\in \lbrace g,B,W \rbrace$, respectively. The ALP 
mass $m_a$ and the scale $f_a$ are assumed to be independent parameters, in contrast to the QCD-axion paradigm, which 
is characterized by the relation $m_a \, f_a \approx m_\pi \, f_\pi$. 

At the energy-scales relevant at $B$-factories, the ALP interactions with the $Z$ boson can be safely neglected, due to the 
Fermi constant suppression. Furthermore, the  ALP couplings to the top-quark and $W^\pm$ boson are relevant only to the 
study of flavor-changing neutral currents observables, which are complementary to the probes discussed here. The only 
relevant couplings in Eq.~\eqref{eq:leffalp0} at low-energies, for our purposes, are then:
%%%%%%%%%%%%%%%%
\begin{align}
\begin{split}
\delta\mathcal{L}_{\mathrm{eff}} \, \supset \, &\dfrac{1}{2} (\partial^\mu a) \, (\partial_\mu a) - \dfrac{m_a^2}{2}a^2 \\
&-    \dfrac{c_{a\gamma\gamma}}{4 } \dfrac{a}{f_a} F_{\mu\nu}\widetilde{F}_{\mu\nu} 
  - \dfrac{c_{agg}}{4} \dfrac{a}{f_a}\, G_{a}^{\mu\nu} \widetilde{G}^a_{\mu\nu} - \dfrac{\partial_\mu a}{2 f_a} 
    \sum_f c_{aff}\, \overline{f}\gamma^\mu \gamma_5 f\,,
\end{split}
\label{eq:leff-alp}
\end{align}
%%%%%%%%%%%%%%%
where $c_{a\gamma\gamma} = c_{aBB}\, \cos^2 \theta_W + c_{aWW}\, \sin^2 \theta_W$. The couplings relevant to ALP 
production at B-factories are $\lbrace c_{a\gamma\gamma}, c_{abb} \rbrace$, while the other couplings only contribute 
to the ALP branching fractions. 

Light pseudoscalar particles can also act as portals to a light dark sector. In this case, to describe these additional 
interactions, new couplings are customarily introduced. By assuming, for instance, an extra light and neutral dark fermion 
state $\chi$, the following term should be considered in the effective Lagrangian:
\bea
 \delta\mathcal{L}_{\mathrm{eff}} \, \supset \, - c_{a\chi\chi}\,\dfrac{\partial_\mu a}{2 f_a} \,
 \overline{\chi}\gamma^\mu \gamma_5 \chi\,,
\eea
where $c_{a\chi\chi}$ denotes a generic coupling, which can induce a sizable ALP decay into invisible final states, 
as will be considered in the following. 

\subsection{$B$-factories probes of invisible ALPs}
\label{subsec:BFACT}

%%%%%%%%%%%%%%%%%%%%%%%%%%%%%%%%%%%%%%%%%%%%%%%%%%
%\section{Invisible ALPS}
%%%%%%%%%%%%%%%%%%%%%%%%%%%%%%%%%%%%%%%%%%%%%%%%%%

The main goal of this talk will be to revisit the theoretical expressions available in the literature, including 
ALP coupling to bottom quarks\footnote{The case of ALP coupling to charm quarks will be shortly discussed at the 
end of the talk.}, as well as previously unaccounted experimental uncertainties~\cite{Merlo:2019anv}.
In the following, the \textit{invisible ALP} scenario will be considered, i.e the scenario in which the coupling 
to the dark sector $c_{a\chi\chi}$ is large, in comparison to the SM couplings, and therefore the ALP will decay 
predominantly into an invisible channel, providing the mono-$\gamma$ plus missing energy signature. Even though 
the main focus will be the minimalistic invisible ALP scenario, most of the observations can be translated mutatis 
mutandis to the visible case.

\begin{figure}[t]
\centering
\includegraphics[width=.4\linewidth]{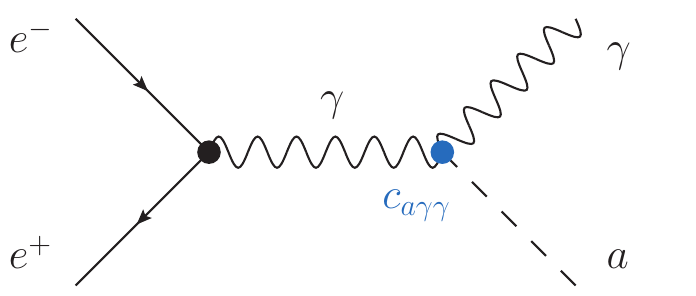}
\caption{\small \em Non-resonant contribution to the process $e^+ e^-\to \gamma a$ produced via the effective 
coupling $c_{a\gamma\gamma}$ defined in Eq.~\eqref{eq:leff-alp}.}
\label{fig:diag-non-resonant}
\end{figure}

\paragraph*{\bf Non-resonant ALP Production.} 
The most straightforward way of producing ALPs in $e^+e^-$ facilities is via the non-resonant process $e^+ e^-\to\gamma a$, 
as illustrated in Fig.~\ref{fig:diag-non-resonant}. If the ALP does not decay inside the detector, or decay invisibly, 
this process would result in an energetic $\gamma$ plus missing energy. The total cross-section, keeping explicit the 
ALP mass dependence, is
\bea
\label{eq:sigma-monogamma}
\sigma_{\text{NR}} (s) & = & \dfrac{\alpha_{\mathrm{em}}}{24} \dfrac{c_{a\gamma\gamma}^2}{f_a^2}
\left(1- \dfrac{m_a^2}{s}\right)^3\,.
\eea
The contributions coming from the exchange of an off-shell $Z$ boson, which are also induced by $c_{aWW}$ in 
Eq.~\eqref{eq:leffalp0}, have been neglected, since they are suppressed, at low-energies, by $s/m_Z^2 \ll 1$. 

While the non-resonant contribution to ALP production given above is unavoidable in any experiment relying on $e^+ e^-$ 
collisions, the situation at $B$-factories is more intricate since these experiments operate at specific $\Upsilon(nS)$ 
resonances. Therefore, it is crucial to account for the resonantly enhanced contributions, which can be numerically 
significant. 

%%%%%%%%%%%%%%%%%%%%%%%%%%%%%%%%%%%%%%%%
%%%%%%%%%%%%%%%%%%%%%%%%%%%%%%%%%%%%%%%%

\begin{figure*}[!h]
\centering
\includegraphics[scale=.88]{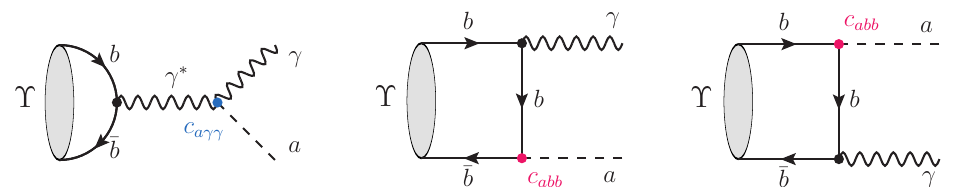}
\caption{\small \em Contributions to the $\Upsilon(nS)\to \gamma a$ decays from the effective couplings introduced in 
the Lagrangian or Eq.~(\eqref{eq:leff-alp}).}
\label{fig:diag-resonant}
\end{figure*}

\paragraph*{\bf Resonant ALP Production.}
Vector quarkonia can produce significant resonant contributions to the mono-$\gamma$ channel, $e^+e^-\to\Upsilon\to\gamma a$, 
since they are very narrow particles coupled to the electromagnetic current. Assuming a fixed center-of-mass energy 
$\sqrt{s} \approx m_\Upsilon$, as is the case at $B$-factories, and using the Breit-Wigner approximation, one finds for 
the resonant cross-section 
\begin{equation}
\label{eq:res-xsection}
\sigma_\mathrm{R}(s) = \sigma_{\mathrm{peak}}\,\dfrac{ m_\Upsilon^2\Gamma_\Upsilon^2}{(s-m_\Upsilon^2)^2+m_\Upsilon^2 
\Gamma_\Upsilon^2}\,\mathcal{B}(\Upsilon\to \gamma a)\,, 
\end{equation}
where $m_\Upsilon$ and $\Gamma_\Upsilon$ are the mass and width of a specific $\Upsilon$ resonance, and $\sigma_{\mathrm{peak}}$ 
is the peak cross-section defined as
\begin{equation}
\sigma_{\mathrm{peak}} = \dfrac{12 \pi \mathcal{B}(\Upsilon \to ee)}{m_\Upsilon^2}\,,
\end{equation}
with $\mathcal{B}(\Upsilon \to ee)$ being the leptonic branching fraction, experimentally determined for the different 
$\Upsilon(nS)$ resonances. The effective couplings defined in Eq.~\eqref{eq:leff-alp} appear, instead, in the 
$\mathcal{B}(\Upsilon\to \gamma a)$ branching fraction, as illustrated in Fig.~\ref{fig:diag-resonant}, that reads 
\cite{Merlo:2019anv}:
\bea
\label{eq:ups-general}
\mathcal{B}(\Upsilon \to \gamma a) & = &\dfrac{\alpha_\mathrm{em}}{216\,\Gamma_\Upsilon} {m_\Upsilon f_\Upsilon^2 }
\left(1-\dfrac{m_a^2}{m_\Upsilon^2}\right) \, \Bigg{[} \dfrac{c_{a\gamma\gamma}}{f_a} 
\left(1-\dfrac{m_a^2}{m_\Upsilon^2}\right)- 2\dfrac{c_{abb}}{f_a} \Bigg{]}^2\,.
\eea
Note, however, that here the computation of the $c_{abb}$ contributions are done within a first approximation that simplifies 
the QCD structure-dependent emission of this decay \cite{Guerrera:2021yss,Guerrera:2022ykl}. If a NP signal is indeed observed, 
a more accurate theoretical calculation would be needed to fully assess the (non-perturbative) effects associated to the last 
two diagrams in Fig.~\ref{fig:diag-resonant}.

As shown in Eq.~\eqref{eq:ups-general}, the $c_{a\gamma\gamma}$ and $c_{abb}$ couplings can induce comparable contributions 
to the resonant cross-section in Eq.~\eqref{eq:res-xsection}. Moreover, depending on the relative sign, these two 
couplings can interfere destructively or constructively. The previous discussion implies that the resonant contributions 
are not only important to correctly assess limits on the ALP coupling to photons, $c_{a\gamma\gamma}$, but they also open 
the window to probe the ALP coupling to $b$-quarks, $c_{abb}$. Finally, note that Eq.~\eqref{eq:ups-general} shows a 
different dependence on $m_a$ and $\lbrace c_{a\gamma\gamma}, c_{abb} \rbrace$ than the non-resonant cross-section 
in Eq.~\eqref{eq:sigma-monogamma}.

%%%%%%%%%%%%%%%%%%%%%%%%%%%%%%%%%%%%%%%%
%%%%%%%%%%%%%%%%%%%%%%%%%%%%%%%%%%%%%%%%
%%%%%%%%%%%%%%%%%%%%%%%%%%%%%%%%%%%%%%%%
%%%%%%%%%%%%%%%%%%%%%%%%%%%%%%%%%%%%%%%%

%
\paragraph*{\bf Non-resonant vs Resonant ALP Production.}
Naively, one would expect that the resonant cross-section {(Eq.~\eqref{eq:res-xsection})} clearly dominates over the 
non-resonant one {(Eq.~\eqref{eq:sigma-monogamma}) for the very narrow $\Upsilon(1S)$, $\Upsilon(2S)$ and $\Upsilon(3S)$ 
resonances, since $\Gamma_\Upsilon/m_\Upsilon \ll 1$. Nevertheless, this turns out to not be always correct in some of 
the cases under consideration, since the resonance widths are typically much narrower than the energy beam resolution, 
$\sigma_W$. Therefore, when the resonance is not fully resolved by the experiment, a convolution of the theoretical 
resonant contribution with a Gaussian spread function, with spread $\sigma_W$, has to be performed (see for 
example~\cite{Eidelman:2016aih}) and an ``experimental'' resonant cross section can be defined as
% since these experiments are intrinsically limited by the energy spread of the $e^+e^-$ beam, which is of order 
% $\sigma_W \approx 5$~MeV at current facilities. This value is considerably larger than 
% the width of these resonances, which therefore cannot be fully resolved at $B$-factories. The only exception is the 
% $\Upsilon(4S)$ resonance, for which $\Gamma_{\Upsilon(4S)}=20.5$~MeV. Therefore, one should expect a sizable reduction 
% of the estimation in Eq.~\eqref{eq:res-xsection} for the lightest quarkonia resonances, due to this intrinsic 
% experimental uncertainty. To account for the beam-energy uncertainties, the procedure presented in 
%  has been adopted by performing a convolution of $\sigma_{\mathrm{R}}(s)$ with a 
% Gaussian distribution, with spread $\sigma_W$,
\begin{equation}
\label{eq:prod-vis}
\langle \sigma_\mathrm{R} (s) \rangle_\mathrm{exp} = \int \! \! \mathrm{d}q \, 
\dfrac{\sigma_\mathrm{R} (q^2)}{\sqrt{2\pi}\sigma_W} 
\mathrm{exp}\left[-\dfrac{(q-\sqrt{s})^2}{2\sigma_W^2}\right]\,.
\end{equation}
At the very narrow $\Upsilon(nS)$ resonances, with $n=1,2,3$, one finds $\Gamma_\Upsilon\ll \sigma_W$, in such a way 
that the previous expression can be simplified by writing:
\begin{equation}
\label{eq:prod-ups}
\langle\sigma_\mathrm{R} (m_\Upsilon^2)\rangle_\mathrm{exp}=\rho\,\sigma_{\mathrm{peak}}\,\mathcal{B}(\Upsilon\to\gamma a)\,,
\end{equation}
where the parameter $\rho$, defined as
\begin{equation}
\rho=\sqrt{\dfrac{\pi}{8}}\dfrac{\Gamma_\Upsilon}{\sigma_W}\,,
\end{equation}
accounts for the cross-section suppression at the peak due to the finite beam-energy spread. 

% %%%%%%%%%%%%
% \subsection{The general case: $c_{a\gamma\gamma}\neq 0$ and $c_{abb}\neq 0$}
% \label{sec:fomulas-2}
% %%%%%%%%%%%%

In Table~\ref{tab:xsection}, the estimation of the resonant and non-resonant cross-section, for each $\Upsilon$ resonance, 
along with the peak cross-section $\sigma_{\mathrm{peak}}$ and the suppression parameter $\rho$. This computation has been 
performed with the Belle-II (KEKB) energy-spread for illustration. From this table, one learns that even though the peak 
cross-section is large for the $\Upsilon(nS)$ resonances ($n=1,2,3$), the beam-energy uncertainties entail a considerable 
suppression of the \textit{experimental} cross-section. These effects are milder for the $\Upsilon(4S)$ resonance, but in turn 
the cross-section at the peak is much smaller in this case. The final results are summarized in the last column of 
Table~\ref{tab:xsection}, which shows that the effective resonant cross-section is smaller than the non-resonant one, 
but it still contributes with numerically significant effects. For the (very) narrow resonances $\Upsilon(nS)$ ($n=1,2,3$), 
the resonant contribution amounts to corrections between $20\%$ and $50\%$ to the non-resonant one, which should be 
included when reinterpreting experimental searches.~\footnote{Interference effects between the non-resonant and resonant 
$c_{a\gamma\gamma}$ terms turn out to be negligible due to the small width of the $\Upsilon(nS)$ resonances.}  
On the other hand, for the $\Upsilon(4S)$ resonance the resonant contribution turns out to be negligible, due to its 
larger width, as expected. 

% These effects will be quantified in the following in two scenarios: (i) ALP with predominant couplings to photons, 
% $|c_{a\gamma \gamma}|\gg |c_{abb}|$, and (ii) the general case with both $c_{a\gamma\gamma}$ and $c_{abb}$ nonzero.

%%%%%%%%%%%%%%%%%%%%%%%%%%%%%%%%%%%%%%
\begin{table*}[!t]
\centering
  \renewcommand{\arraystretch}{1.4} 
  %\footnotesize
\begin{tabular}{|c|cc|cc||c|}
\hline
$\Upsilon(nS)$  & $m_\Upsilon~[\mathrm{GeV}]$ &  $\Gamma_\Upsilon~[\mathrm{keV}]$  & 
$\sigma_{\mathrm{peak}}$~[nb] & $\rho$  & $\langle\sigma_{\mathrm{R}}(m_\Upsilon^2)\rangle_{\mathrm{exp}}/\sigma_{\text{NR}}$\\
\hline\hline
$\Upsilon(1S)$  & $9.460$  & $54.02$  & $3.9(18)\times 10^{3}$             & $6.1\times 10^{-3}$ &  $0.53(5)$ \\
$\Upsilon(2S)$  & $10.023$ &  $31.98$  & $2.8(2)\times 10^{3}$             & $3.7\times 10^{-3}$ &  $0.21(3)$ \\
$\Upsilon(3S)$  & $10.355$ & $20.32$  & $3.0(3)\times 10^{3}$              & $2.3\times 10^{-3}$ &  $0.16(3)$ \\ 
$\Upsilon(4S)$  & $10.580$ & $20.5\times 10^3$  &  $2.10(10)$ & $0.83$    & $3.0(3)\times 10^{-5}$ \\
\hline
\end{tabular}
\caption{\small \em Estimated \textit{visible} cross-section at Belle-II for $e^+e^-\to \Upsilon \to \gamma a$ compared to the 
non-resonant one, $e^+e^-\to \gamma^\ast \to \gamma a$. Here, vanishing ALP couplings with $b$-quarks have been assumed, 
$c_{abb} =0$. }  
\label{tab:xsection}
\end{table*}
%%%%%%%%%%%%%%%%%%%%%%%%%%%%%%%%%%%%%%

%%%%%%%%%%%%%%%%%%%%%%%%%%%%%%%%%%%%%%%%%%%%%%%%%%%%%%%%%%%%
\section{Summary of experimental searches}
\label{sec:exp}

Based on previous observations, ALP searches at $B$-factories can be classified in the following three categories:

%%%%%%%%%%%%%%%
{\bf i) Resonant searches}: Excited quarkonia states $\Upsilon (nS)$ (with $n>1$) can decay into lighter $\Upsilon (nS)$ 
resonances via pion emission, as for example $\Upsilon (2S) \to \Upsilon(1S)\,\pi^+ \pi^-$ and $\Upsilon (3S) \to 
\Upsilon(1S)\,\pi^+ \pi^-$. By exploiting the process kinematics one can reconstruct the $\Upsilon(1S)$ meson and study 
its decay into a specific final state, which can, for instance, be the invisible $\Upsilon$ decay or the $\Upsilon$ decay 
into photon and a light (pseudo)scalar particle. These searches are dubbed \textit{resonant}, since they allow to directly 
probe $\mathcal{B}(\Upsilon\to \gamma a)$ in a \textit{model-independent} way, regardless of the non-resonant contribution 
from Fig.~\ref{fig:diag-non-resonant}. In other words, reported limits on $\mathcal{B}
(\Upsilon(1S)\to\gamma a)$ can be used to constrain both $c_{a\gamma\gamma}$ and $c_{abb}$ via Eq.~\eqref{eq:ups-general}. 
Searches along these lines have been performed, for instance, by BaBar \cite{delAmoSanchez:2010ac} and, more recently, 
by Belle~\cite{Seong:2018gut}, under the assumption that the ALP does not decay into visible particles inside the detector.

{\bf ii) Mixed searches}: Alternatively, experimental searches could be performed at $\Upsilon(nS)$ (with $n=1,2,3$) without 
identifying the $\Upsilon$ decay from a secondary vertex. Example of such experimental searches are the ones~\cite{Aubert:2008as} 
performed at $\sqrt{s}=m_{\Upsilon(3S)}$, where limits on $\mathcal{B}(\Upsilon(3S)\to\gamma a)\times\mathcal{B}(a\to\mathrm{inv})$ 
are extracted from the total $e^+e^-\to\gamma a (\to\mathrm{inv})$ cross-section. From the above discussion, however, it is clear 
that this method is probing both resonant (Eq.~\eqref{eq:prod-ups}) and non-resonant (Eq.~\eqref{eq:sigma-monogamma}) cross-sections 
and therefore model-independent limits on $\mathcal{B}(\Upsilon (3S)\to \gamma a)$ could not be extracted from these experimental 
results. The only scenarios for which such limits can be derived are the ones with $|c_{a\gamma\gamma}| \ll |c_{abb}|$, as predicted 
in models with an extended Higgs sector, since the non-resonant cross-section practically vanishes in this case. In the most 
general ALP scenario, instead, the limits on $\lbrace c_{a\gamma\gamma}, c_{abb} \rbrace$ can be obtained from via a rescaling factor,
%%%%%%%%%%%%%%%%
\begin{align}
\label{eq:xsection-total}
\begin{split}
\dfrac{\langle\sigma_{\text{R}}(s) + \sigma_{\text{NR}}(s)  \rangle_{\mathrm{vis}}}{\langle\sigma_{\mathrm{R}} \rangle_{\mathrm{vis}}} 
&\approx 1+\dfrac{\sigma_{\text{NR}}}{\langle\sigma_{\mathrm{R}}\rangle_\mathrm{vis}} \,,
\end{split}
\end{align}
%%%%%%%%%%%%%%%%
which accounts for the non-resonant contributions (Eq.~\eqref{eq:prod-vis}) that have been overlooked experimentally in the 
total cross-section. Note, also, that similar effects have also been overlooked in reinterpretations of other experimental 
limits, as for example the ones on $\mathcal{B}(\Upsilon(3S)\to \gamma a)\times \mathcal{B}(a\to \mathrm{hadrons})$ to 
constrain the product of ALP couplings to photons and gluons. \footnote{The reinterpretation described above has a possible 
caveat related to the treatment of the background. See~\cite{Merlo:2019anv} for a detailed discussion on this point.}
 
{\bf iii) Non-resonant searches}: The resonant cross-section is negligible at the $\Upsilon(4S)$ resonance, as can be 
seen from Table~\ref{tab:xsection}, since its mass lies just above the $B \overline{B}$ production threshold. Therefore, 
experimental searches at the $\Upsilon(4S)$ resonance can only probe the $c_{a\gamma\gamma}$ coupling via the non-resonant 
ALP production illustrated in Fig.~\ref{fig:diag-non-resonant}. To our knowledge, no such ALP {\it invisible} search has 
been performed yet at $B$-factories. For an order of magnitude estimation one can recast the $\mathcal{B}(\Upsilon(4S)\to\gamma a)$ 
limits extracted from Belle II data from $\Upsilon(4S)\to\gamma\,a, (a\to\gamma\gamma)$ decay \cite{Belle-II:2020jti}.

\section{Constraining the ALP parameter space}
\label{sec:constraints}

Constraints on the $\lbrace c_{a\gamma\gamma},c_{abb}\rbrace/f_a$ space are shown in Fig.~\ref{fig:constraints-alps-2param}, 
considering, for illustration, the \emph{invisible} ALP scenario, i.e. by assuming that $\mathcal{B}(a\to \mathrm{inv})=1$, 
or equivalently, that the ALP does not decay inside the detector. Clearly, the results derived below can be easily recast 
to scenarios in which the invisible ALP branching fraction is smaller than one, see for example~\cite{DiLuzio:2024jip}. 

The Left (Right) plot refers to the $c_{abb}/c_{a\gamma\gamma}>0$ ($c_{abb}/c_{a\gamma\gamma}<0$) sign choice, for the chosen 
reference value of $m_a=0.2$~GeV. As expected from Eq.~\eqref{eq:ups-general} for the $c_{abb}/c_{a\gamma\gamma}>0$, 
a destructive interference between the fermion and photon ALP coupling can appear. In this case, the (resonant) $\Upsilon(1S)$ 
constraints have a flat direction that cannot be resolved by only relying on resonant ALP searches. The combination of 
couplings that lead to this cancellation depends on the ALP mass, especially for $m_a$ values near the kinematical 
threshold. BaBar results obtained at the $\Upsilon(3S)$ resonance, which is not reconstructed, show a different sensitivity 
to $\lbrace c_{a\gamma\gamma},c_{abb}\rbrace$, (blue region). While a cancellation between $c_{a\gamma\gamma}$ and $c_{abb}$ 
is possible for resonant cross-section, this cannot occur for the non-resonant one, which depends only on the $c_{a\gamma\gamma}$ 
coupling. The combination of these complementary searches allows one to corner the ALP parameter space as depicted in 
Fig.~\ref{fig:constraints-alps-2param}. Moreover, the recast of the 3$\gamma$ search performed at Belle II, operating at the 
$\Upsilon(4S)$ resonance are displayed in the same plot (black line). Finally, in 
Fig.~\ref{fig:constraints-alps-2param} the available information on the invisible ALP scenario from radiatice $J/\Psi$ decays 
in the $\lbrace g_{a\gamma\gamma} , g_{acc}\rbrace$ parameter space, are shown for comparison. The dark and light red areas show the 
exclusion regions from the invisible~\cite{BESIII:2020sdo} and the 3$\gamma$ recast~\cite{BESIII:2022rzz} BESIII analyses.
As one can see, B-Factory and Charm-Factory sensitivities on the corresponding $\lbrace g_{a\gamma\gamma} , g_{aQQ}\rbrace$ 
parameter space are comparable.

%%%%%%%%%%%%%%%%%%%%%%%%%%%%%%%%%%%%%%%%
%%%%%%%%%%%%%%%%%%%%%%%%%%%%%%%%%%%%%%%%
\begin{figure*}[t!]
\centering
\includegraphics[width=0.95\linewidth]{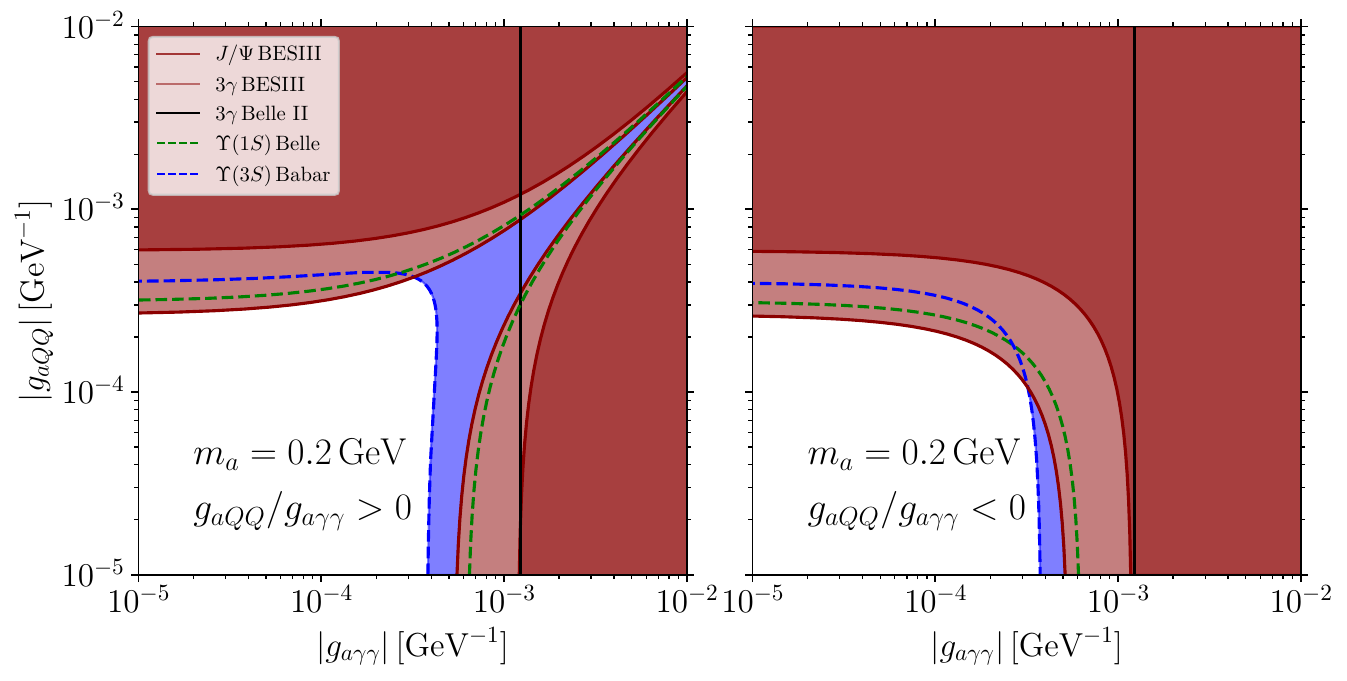}
\caption{\small \em Excluded $\{g_{a\gamma\gamma},g_{a f\!f}\}$ parameter space for the {\it invisible} scenario. Dashed-line constraints 
are taken from~\cite{Merlo:2019anv}. The red (light-red) area is obtained from BESIII invisible search~\cite{BESIII:2020sdo} (BESIII 3$\gamma$ 
recast search \cite{BESIII:2022rzz}) while the black line represents Belle II limit from the 
3$\gamma$ (recast) search~\cite{Belle-II:2020jti}.
Resonant radiative decays of $\Upsilon$ and $J/\psi$ test different fermionic parameters, $g_{aQQ}=g_{abb}$ or $g_{acc}$ respectively.}
\label{fig:constraints-alps-2param}
\end{figure*}
%%%%%%%%%%%%%%%%%%%%%%%%%%%%%%%%%%%%%%%%
%%%%%%%%%%%%%%%%%%%%%%%%%%%%%%%%%%%%%%%%

\section*{Acknowledgments}

This work received funding from the European Union's Horizon 2020 research and innovation programme under the Marie 
Sk\l{}odowska-Curie grant agreement N$^{\circ}$ 860881-HIDDeN, grant agreement N$^{\circ}$ 101086085–ASYMMETRY 
and by the INFN Iniziative Specifica APINE. This work was also partially supported by the Italian MUR Departments of 
Excellence grant 2023-2027 ``Quantum Frontiers''. The work of LDL is also supported by the project ``CPV-Axion'' under 
the Supporting TAlent in ReSearch@University of Padova (STARS@UNIPD).


\section*{References}

\begin{thebibliography}{99}
%\cite{Peccei:1977hh}
\bibitem{Peccei:1977hh} 
  R.~D.~Peccei and H.~R.~Quinn,
  %``CP Conservation in the Presence of Instantons,''
  Phys.\ Rev.\ Lett.\  {\bf 38}, 1440 (1977).
  %doi:10.1103/PhysRevLett.38.1440
  %%CITATION = doi:10.1103/PhysRevLett.38.1440;%%
  %4714 citations counted in INSPIRE as of 06 May 2019


%\cite{Wilczek:1977pj}
\bibitem{Wilczek:1977pj} 
  F.~Wilczek,
  %``Problem of Strong  $P$  and  $T$  Invariance in the Presence of Instantons,''
  Phys.\ Rev.\ Lett.\  {\bf 40}, 279 (1978).
  %doi:10.1103/PhysRevLett.40.279
  %%CITATION = doi:10.1103/PhysRevLett.40.279;%%
  %3148 citations counted in INSPIRE as of 06 May 2019


%\cite{Weinberg:1977ma}
\bibitem{Weinberg:1977ma} 
  S.~Weinberg,
  %``A New Light Boson?,''
  Phys.\ Rev.\ Lett.\  {\bf 40}, 223 (1978).
  %doi:10.1103/PhysRevLett.40.223
  %%CITATION = doi:10.1103/PhysRevLett.40.223;%%
  %3280 citations counted in INSPIRE as of 06 May 2019


%\cite{Kim:1979if}
\bibitem{Kim:1979if} 
  J.~E.~Kim,
  %``Weak Interaction Singlet and Strong CP Invariance,''
  Phys.\ Rev.\ Lett.\  {\bf 43}, 103 (1979).
  %doi:10.1103/PhysRevLett.43.103
  %%CITATION = doi:10.1103/PhysRevLett.43.103;%%
  %1949 citations counted in INSPIRE as of 06 May 2019


%\cite{Shifman:1979if}
\bibitem{Shifman:1979if} 
M.~A.~Shifman, A.~I.~Vainshtein and V.~I.~Zakharov,
%``Can Confinement Ensure Natural CP Invariance of Strong Interactions?,''
Nucl.\ Phys.\ B {\bf 166}, 493 (1980).
%doi:10.1016/0550-3213(80)90209-6
%%CITATION = doi:10.1016/0550-3213(80)90209-6;%%
%1680 citations counted in INSPIRE as of 06 May 2019


%\cite{Zhitnitsky:1980tq}
\bibitem{Zhitnitsky:1980tq} 
A.~R.~Zhitnitsky,
%``On Possible Suppression of the Axion Hadron Interactions. (In Russian),''
Sov.\ J.\ Nucl.\ Phys.\  {\bf 31}, 260 (1980)
[Yad.\ Fiz.\  {\bf 31}, 497 (1980)].
%%CITATION = SJNCA,31,260;%%
%1326 citations counted in INSPIRE as of 06 May 2019


%\cite{Dine:1981rt}
\bibitem{Dine:1981rt} 
M.~Dine, W.~Fischler and M.~Srednicki,
%``A Simple Solution to the Strong CP Problem with a Harmless Axion,''
Phys.\ Lett.\  {\bf 104B}, 199 (1981).
%doi:10.1016/0370-2693(81)90590-6
%%CITATION = doi:10.1016/0370-2693(81)90590-6;
%%%2372 citations counted in INSPIRE as of 06 May 2019

%\cite{Merlo:2019anv}
\bibitem{Merlo:2019anv}
L.~Merlo, F.~Pobbe, S.~Rigolin and O.~Sumensari,
%``Revisiting the production of ALPs at B-factories,''
JHEP \textbf{06} (2019), 091.
%doi:10.1007/JHEP06(2019)091
%[arXiv:1905.03259 [hep-ph]].
%28 citations counted in INSPIRE as of 11 May 2023

%\cite{DiLuzio:2024jip}
\bibitem{DiLuzio:2024jip}
L.~Di Luzio, A.~W.~M.~Guerrera, X.~Ponce D\'\i{}az and S.~Rigolin,
%``Axion-Like Particles in Radiative Quarkonia Decays,''
[arXiv:2402.12454 [hep-ph]].
%0 citations counted in INSPIRE as of 22 Feb 2024

%\cite{Brivio:2017ije}
\bibitem{Brivio:2017ije} 
  I.~Brivio, M.~B.~Gavela, L.~Merlo, K.~Mimasu, J.~M.~No, R.~del Rey and V.~Sanz,
  %``ALPs Effective Field Theory and Collider Signatures,''
  Eur.\ Phys.\ J.\ C {\bf 77}, no. 8, 572 (2017).
  %doi:10.1140/epjc/s10052-017-5111-3
  %[arXiv:1701.05379 [hep-ph]].
  %%CITATION = doi:10.1140/epjc/s10052-017-5111-3;%%
  %42 citations counted in INSPIRE as of 06 May 2019

%\cite{Eidelman:2016aih}
\bibitem{Eidelman:2016aih} 
  S.~Eidelman, D.~Epifanov, M.~Fael, L.~Mercolli and M.~Passera,
  %``$\tau$ dipole moments via radiative leptonic $\tau$ decays,''
  JHEP {\bf 1603}, 140 (2016).
%  doi:10.1007/JHEP03(2016)140
%  [arXiv:1601.07987 [hep-ph]].
  %%CITATION = doi:10.1007/JHEP03(2016)140;%%
  %23 citations counted in INSPIRE as of 06 May 2019

%\cite{Guerrera:2021yss}
\bibitem{Guerrera:2021yss}
A.~W.~M.~Guerrera and S.~Rigolin,
%``Revisiting $K \rightarrow \pi a$ decays,''
Eur. Phys. J. C \textbf{82} (2022) no.3, 192.
%doi:10.1140/epjc/s10052-022-10146-x
%[arXiv:2106.05910 [hep-ph]].
%10 citations counted in INSPIRE as of 14 May 2023

%\cite{Guerrera:2022ykl}
\bibitem{Guerrera:2022ykl}
A.~W.~M.~Guerrera and S.~Rigolin,
%``ALP Production in Weak Mesonic Decays,''
Fortsch. Phys. \textbf{71} (2023) no.2-3, 2200192.
%doi:10.1002/prop.202200192
%[arXiv:2211.08343 [hep-ph]].
%3 citations counted in INSPIRE as of 14 May 2023

%\cite{delAmoSanchez:2010ac}
\bibitem{delAmoSanchez:2010ac} 
  P.~del Amo Sanchez {\it et al.} [BaBar Collaboration],
  %``Search for Production of Invisible Final States in Single-Photon Decays of $\Upsilon(1S)$,''
  Phys.\ Rev.\ Lett.\  {\bf 107}, 021804 (2011).
%  doi:10.1103/PhysRevLett.107.021804
%  [arXiv:1007.4646 [hep-ex]].
  %%CITATION = doi:10.1103/PhysRevLett.107.021804;%%
  %67 citations counted in INSPIRE as of 06 May 2019

%\cite{Seong:2018gut}
\bibitem{Seong:2018gut} 
  I.~S.~Seong {\it et al.} [Belle Collaboration],
  %``Search for a light $CP$-odd Higgs boson and low-mass dark matter at the Belle experiment,''
  Phys.\ Rev.\ Lett.\  {\bf 122}, no. 1, 011801 (2019).
%  doi:10.1103/PhysRevLett.122.011801
%  [arXiv:1809.05222 [hep-ex]].
  %%CITATION = doi:10.1103/PhysRevLett.122.011801;%%
  %2 citations counted in INSPIRE as of 06 May 2019

%\cite{Aubert:2008as}
\bibitem{Aubert:2008as} 
  B.~Aubert {\it et al.} [BaBar Collaboration],
  %``Search for Invisible Decays of a Light Scalar in Radiative Transitions $\upsilon_{3S} \to \gamma$ A0,''
  arXiv:0808.0017 [hep-ex].
  %%CITATION = ARXIV:0808.0017;%%
  %115 citations counted in INSPIRE as of 06 May 2019

%\cite{Belle-II:2020jti}
\bibitem{Belle-II:2020jti}
F.~Abudin\'en \textit{et al.} [Belle-II],
%``Search for Axion-Like Particles produced in $e^+e^-$ collisions at Belle II,''
Phys. Rev. Lett. \textbf{125} (2020) no.16, 161806.
%doi:10.1103/PhysRevLett.125.161806
%[arXiv:2007.13071 [hep-ex]].
%82 citations counted in INSPIRE as of 12 May 2023

%\cite{BESIII:2020sdo}
\bibitem{BESIII:2020sdo}
M.~Ablikim \textit{et al.} [BESIII],
%``Search for the decay $J/\psi\to\gamma + \rm {invisible}$,''
Phys. Rev. D \textbf{101} (2020) no.11, 112005
doi:10.1103/PhysRevD.101.112005
[arXiv:2003.05594 [hep-ex]].
%14 citations counted in INSPIRE as of 20 Feb 2024

%\cite{BESIII:2022rzz}
\bibitem{BESIII:2022rzz}
M.~Ablikim \textit{et al.} [BESIII],
%``Search for an axion-like particle in radiative J/\ensuremath{\psi} decays,''
Phys. Lett. B \textbf{838} (2023), 137698
doi:10.1016/j.physletb.2023.137698
[arXiv:2211.12699 [hep-ex]].
%13 citations counted in INSPIRE as of 20 Feb 2024

\end{thebibliography}
\end{document}